\documentclass[%
 reprint,
superscriptaddress,
 amsmath,amssymb,
 prx,
]{revtex4-2}

\usepackage{ulem}
\usepackage{graphicx}
\usepackage{color} 
\usepackage{dcolumn}
\usepackage{bm}
\usepackage{amsmath}
\usepackage{dashrule}

\begin{document}



\title{Spatial organization of phase-separated DNA droplets}%

\author{Sam Wilken}
\affiliation{%
Physics Department, University of California Santa Barbara, Santa Barbara, CA 93106  USA
}%
\affiliation{%
Materials Department, University of California Santa Barbara, Santa Barbara, CA 93106  USA
}%
\author{Aria Chaderjian}
\affiliation{%
Physics Department, University of California Santa Barbara, Santa Barbara, CA 93106  USA
}%
\author{Omar A. Saleh}
\affiliation{%
Physics Department, University of California Santa Barbara, Santa Barbara, CA 93106  USA
}%
\affiliation{%
Materials Department, University of California Santa Barbara, Santa Barbara, CA 93106  USA
}%

\date{\today}

\begin{abstract}

Many recent studies of liquid-liquid phase separation in biology focus on phase separation as a dynamic control mechanism for cellular function, but it can also result in complex mesoscopic structures.
We primarily investigate a model system consisting of DNA nanostars: finite-valence, self-assembled particles that form micron-scale liquid droplets via a binodal phase transition.
We demonstrate that, upon phase separation, nanostar droplets spontaneously form hyperuniform structures, a type of disordered material with `hidden order' that combines the long-range order of crystals with the short-range isotropy of liquids. 
We find that the hyperuniformity of the DNA droplets reflects near-equilibrium dynamics, where phase separation drives the organization of droplets that then relax toward equilibrium via droplet Brownian motion.
We engineer a two-species system of immiscible DNA droplets and find two distinctly hyperuniform structures in the same sample, but with random cross-species droplet correlations, which rules out explanations that rely on droplet-droplet hydrodynamic interactions.
In addition, we perform experiments on the electrostatic coacervation of peptides and nucleotides which exhibit hyperuniform structures indistinguishable from DNA nanostars, indicating the phenomenon generally applies to phase-separating systems that experience Brownian motion.
Our work on near-equilibrium droplet assembly and structure provides a foundation to investigate droplet organizational mechanisms in driven/biological environments.
This approach also provides a clear path to implement phase-separated droplet patterns as exotic optical or  mechanical metamaterials, or as efficient biochemical reactors.

\end{abstract}

\maketitle

\section{Introduction}

Liquid-liquid phase transitions represent an integral and pervasive mechanism underlying cellular organization \cite{hymanLiquidLiquidPhaseSeparation2014,albertiConsiderationsChallengesStudying2019}.
Most recent work on such phase-separated condensates has focused on their local biomolecular organization (droplet material properties, composition and phase behavior~\cite{shinLiquidPhaseCondensation2017}), or on the regulatory effects they have on the cell~\cite{o2021role}. 
In contrast, only a few studies have investigated long length-scale organization in phase-separating systems, including a theoretical investigation of the Cahn-Hilliard equation~\cite{maRandomScalarFields2017}, and an experiment that posited phase separation forms the material structures that create non-iridescent structural colors of bird feathers~\cite{saranathanStructureOpticalFunction2012}, a phenomenon typically associated with long-range order.
Whether there exists general structural features that span all phase-separating systems remains a profound and yet unexplored problem with potential implications for the material properties of such systems as well as the reaction/regulation effects on cellular function.




The description of long-range material structure frequently focuses on the limits of perfect (i.e.~crystalline) order, or complete disorder (e.g.~as in an ideal gas).
Others exist between those limits, notably `disordered hyperuniform' structures~\cite{torquatoLocalDensityFluctuations2003b}, which are isotropic configurations that are locally disordered (i.e. non-crystalline), but nonetheless display more long-range regularity than expected for an ideal gas.
The characterization of disordered hyperuniform structures have been important to understanding organizational mechanisms for various systems, including the packing of spheres~\cite{wilkenRandomClosePacking2021}, active matter~\cite{zhangHyperuniformActiveChiral2022}, the structure of the early universe~\cite{gabrielliGlasslikeUniverseRealspace2002}, and a universality class of dynamical phase transitions~\cite{hexnerHyperuniformityCriticalAbsorbing2015a}.
Disordered hyperuniform structures also display exotic transport and optical properties, e.g. large isotropic photonic band gaps~\cite{florescu2009designer}, which has resulted in substantial interest in their technological value, for example in fabricating free-form waveguides~\cite{man2013isotropic}.
Hyperuniformity has also been observed in multicomponent systems, like the arrangement of the five types of cones in the chicken eye retina~\cite{jiaoAvianPhotoreceptorPatterns2014}, in which each cone type is individually hyperuniform, and, simultaneously, all cone types together are hyperuniform.
The concurrent hyperuniformity suggests that each component interacts with itself as well as with the other components, though the microscopic mechanism for this is still unclear.
In addition, many of the documented examples of hyperuniformity in experimental systems are observations. Demonstrating control of the parameters that modulate the long-range structure in experiments, especially on the colloidal scale, remains a challenge.

As with other types of long-range order, hyperuniformity is analyzed through the spatial correlations in material density.
Fluctuations in such density correlations are notably suppressed in hyperuniform materials relative to completely random structures.
This can be quantified through the scattering function, $\psi(q)$ (which is the Fourier transform of the density-density correlations). 
Random, ideal-gas configurations show constant values of the scattering function ($\psi(q) \sim q^0$), while hyperuniform structures instead display a low $q$ (long-range) power-law scaling, $\psi(q) \sim q^\alpha$ with $\alpha >0$.  
Hyperuniform scaling has been observed in a few experimental phase-separating systems, including decomposing polymer mixtures~\cite{hashimoto1986late}, with $\alpha = 2$, and solid-state dewetting in semiconductors~\cite{salvalaglio2020hyperuniform}, $\alpha = 4-6$.

To investigate the long-range structures formed by biomolecular phase separation, we perform experiments on DNA nanostar condensates and peptide-nucleotide coacervates.
Our primary focus is on nanostars, a model system that phase-separates to form liquid DNA droplets, are roughly 10 nm, finite-valence, self-assembled DNA particles consisting of four double-stranded ``arms", each decorated with a single-stranded palindromic sticky-end~\cite{biffiPhaseBehaviorCritical2013} (Fig.~\ref{fig:NS_CH}a).
The sticky-ends enable two nanostars to bind through sequence-specific DNA hybridization, with a strength and specificity that is well known and easily modulated by sequence design and the ionic strength of the solvent~\cite{santalucia2004thermodynamics}.
Phase separation takes place when nanostar binding is energetically favorable, creating a percolated-network condensate with liquid-like features due to the transient nature of the nanostar-nanostar bonds (Fig.~\ref{fig:NS_CH}a).
The phase diagram controlling condensation displays features similar to binodal demixing, notably including a coexistence regime where dense liquid droplets are in equilibrium with a dilute nanostar solution~\cite{biffiPhaseBehaviorCritical2013}.
In addition, the sequence specificity of the nanostars provides an avenue to design separate particles with orthogonal attractive interactions that drive multicomponent phase separation~\cite{jeonSequenceControlledAdhesionMicroemulsification2020,sato2021capsule,do2022engineering}.
Finally, we test the generality of the structures formed by LLPS by investigating phase-separated droplets formed by the coacervation of a positively charged polymer (polylysine) with a small negatively-charged molecule (ATP).

We find that phase-separating, nanometer-scale molecules organize into droplet patterns with highly correlated structures on diverging length scales due to an interplay between Cahn-Hilliard spinodal decomposition and droplet Brownian motion.
Both DNA nanostar droplets and polylysine/ATP coacervates, two representative yet distinct phase-separating systems, exhibit quantitatively-identical hyperuniformity, indicating a general mechanism governing phase-separating soft matter systems.
The bottom-up fabrication of self-organized, disordered, hyperuniform structures make these phase-separated soft systems well-suited to various potential applications (e.g. photonic band gap materials, biochemical reactors, and mechanical metamaterials).
Finally, the hyperuniformity that we observe results from a near-equilibrium relaxation and therefore provides a foundation to utilize droplet structural features as a probe of the organizational mechanisms in active/non-equilibrium systems, like those ubiquitous in biology.

\section{Nanostar Droplet Hyperuniformity Mechanism}

\begin{figure*}
\includegraphics[width=\linewidth]{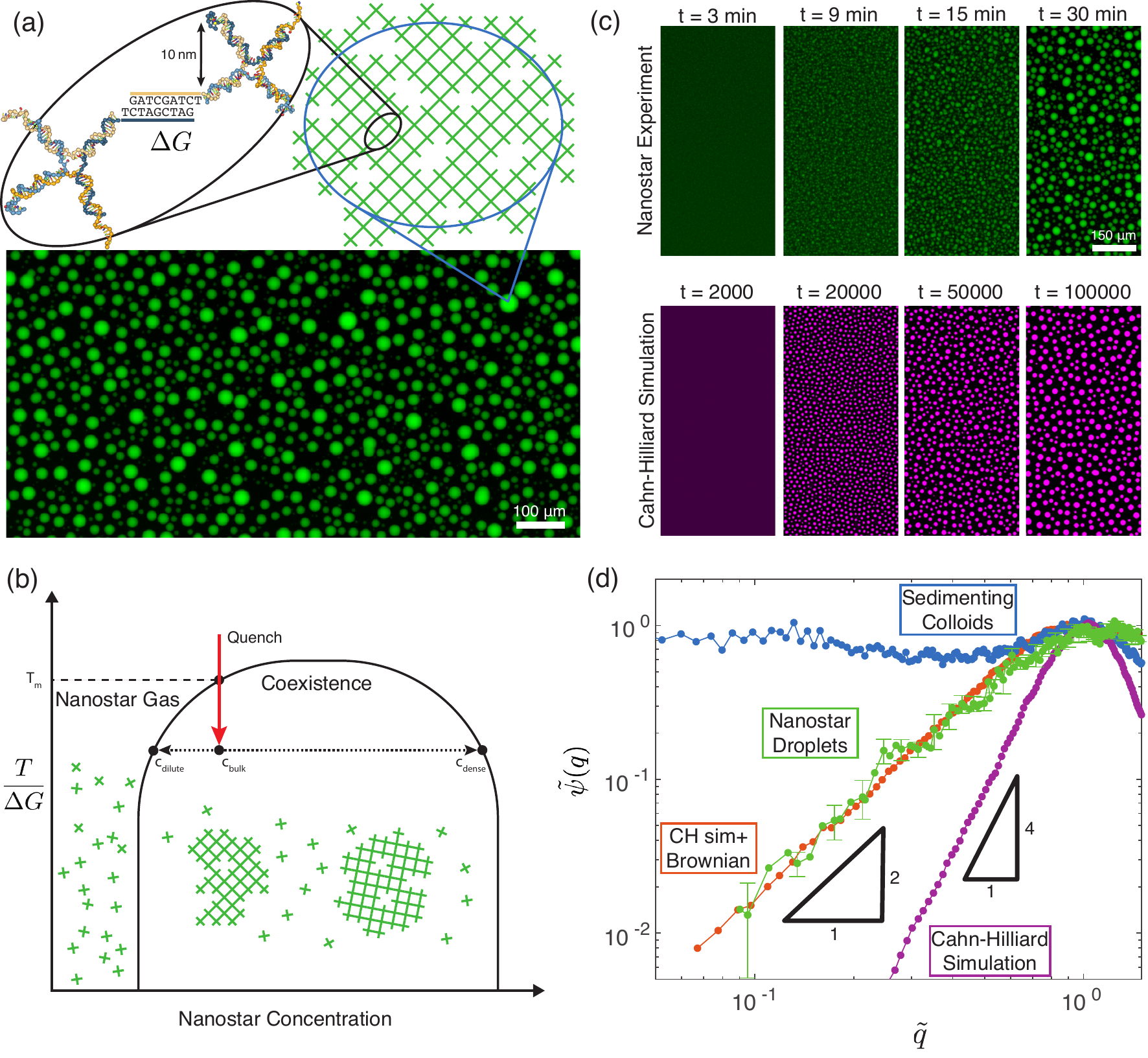}
\caption{\label{fig:NS_CH} Phase-Separating DNA nanostar droplets form hyperuniform structures. (a)~A schematic of the multiscale assembly of nanostar droplets. Nanostars are $\approx$ 10 nm, self-assembled DNA structures that bind, with strength $\Delta G$, through sticky-end interactions, condensing into dynamic, mesh-like networks that form liquid droplets on the micron scale. Bottom: Section of a typical fluorescent image, showing roughly 5,000 condensed nanostar droplets in a 1.2 mm by 0.7 mm area; full experimental images are 1.2 mm by 1.2 mm.
(b)~Schematic of the nanostar phase diagram. Below a critical temperature, nanostars phase separate into droplets with concentration $c_{dense}$ that coexist with unbound, dilute nanostars at concentration, $c_{dilute}$.
(c)~Droplet formation dynamics from experiments on nanostars (top), and from Cahn-Hilliard simulations (bottom). Nanostar solutions are prepared in the melted state $T>T_m$ and quenched at $t=0$. For both the experiment and simulation, droplets form quickly and grow over time, but with characteristically different dynamics.
(d)~The scattering function, $\psi(q)$ of the droplet intensity as a function of wavenumber, $q$, shows hyperuniform scaling $\psi(q \rightarrow 0) \sim q^\alpha$ in the experiment (green), with exponent $\alpha = 2$, as well as the Cahn-Hilliard simulation (magenta), with exponent $\alpha = 4$. 
We recover the experimental structure by randomly displacing Cahn-Hilliard droplets with a displacement magnitude of one droplet spacing (orange).
The scattering function of sedimenting $10\mu m$ diameter polystyrene colloids (blue) do not show hyperuniform structure.
Plotted nanostar droplet experiment $\psi$ is the average of six different quenches and two different sample preparations at a 30 minute quench time. 
Error bars correspond to standard deviation of different quenches and samples. 
Values of $\psi$ are normalized for comparison, $\tilde{\psi}=\psi/\psi_{peak}$ and $\tilde{q}=q/q_{peak}$.}
\end{figure*}

The experimental system consists of tetravalent DNA nanostars, which are formed by mixing together equal amounts of 4 DNA oligos (sequences given in Supplement) that assemble into star-shaped particles with four 20 bp ($\approx 8$ nm) arms tipped with single-stranded palindromic sticky ends (Fig.~\ref{fig:NS_CH}a).
10\% of one strand is modified with a fluorescent dye, Cy5, for imaging. 
Salt (Potassium Acetate, KAce), is added to a fixed concentration of pre-annealed nanostars, to facilitate sticky-end binding. 
The sample is loaded into a microcapillary tube coated with polyacrylamide~\cite{sanchez2013engineering} to prevent droplet adhesion, then visualized on a fluorescent microscope with a temperature control stage.
The sample is heated above the melting temperature ($T_m \approx 45^\circ$C), such that the solution is a homogeneous, single-phase fluid, and quenched into the coexistence regime. 
Liquid DNA droplets form throughout the capillary, then, since the condensate is denser than water~\cite{jeon2018salt}, sediment to the bottom of the capillary where they are imaged (Fig.~\ref{fig:NS_CH}c).

We investigate the structure of the nanostar droplets by calculating the density correlations of the fluorescent images.
In Fourier space, density correlations are revealed by the scattering function, which is defined as $\psi({\bf q}) = \langle \tilde{I}({\bf q}) \tilde{I}(-{\bf q})\rangle$, where $\tilde{I}({\bf q})$ is the discrete Fourier transform of the image $I({\bf r})$ at wavevector ${\bf q}$.
We do not observe anisotropy in the droplet structures (see supplemental material Figure S1), so the scattering function is angularly averaged $\psi({\bf q}) = \psi(|q|) \equiv \psi(q)$.
The low numerical aperture of the objective captures all of the fluorescence of the nanostar droplets (depth of focus$\sim 100~\mu$m) so the two-dimensional images are equivalent to a projection along the optical axis.

For a sample of $30~\mu$M nanostars and $500$ mM KAce, the calculated scattering function shows that the nanostar droplet assemblies are hyperuniform, with a clear reduction of $\psi(q)$ at low $q$.
Particularly, we find a power-law scaling $\psi(q \rightarrow 0)\sim q^\alpha$ with $\alpha = 2.0 \pm 0.1$ on long length scales (Fig.~\ref{fig:NS_CH}d).

The hyperuniformity exponent indicates the organizational principles of the system~\cite{torquato2018hyperuniform}, so investigating which mechanisms produce $\alpha=2$ scaling will provide insight into how phase separating droplet assemblies organize. 
We consider three effects that could modulate the hyperuniform scaling measured in the experiments: sedimentation, phase separation, and Brownian motion.

As nanostar droplets form, they sediment to the bottom surface of the sample chamber so it is possible that droplet-droplet hydrodynamic interactions during sedimentation are responsible for the observed hyperuniform scaling.
Indeed, others have observed $\alpha = 2$ hyperuniform scaling in hydrodynamic calculations of sedimenting colloidal systems, though the suppression of density fluctuations required non-spherically shaped particles~\cite{goldfriend2017screening}.
To test the role of sedimentation independent of phase separation, we perform experiments with  polystyrene spheres of similar size to the droplets (10 $\mu$m diameter) suspended in a 10\% glycerol-water mixture, and at a concentration that, after sedimentation, leads to a 20\% area coverage (similar to the nanostar droplets).
In these conditions, the sedimentation velocity of the colloids roughly matches that of a $10~\mu$m diameter nanostar droplet~\cite{jeon2018salt}, without altering the solvent viscosity significantly. 
We find that the $\psi(q)$ for sedimenting colloids does not display hyperuniformity (Fig.~\ref{fig:NS_CH}d). Instead, the structure looks random ($\psi \sim q^0$).
At intermediate $q$-values, $\psi$ decreases slightly as expected from excluded volume interactions. 
As $q \rightarrow 0$, $\psi$ increases again representing a clustering effect as seen in Rayleigh-Taylor instabilities~\cite{wysocki2009direct}.
We also varied other parameters expected to affect sedimentation, including sample chamber height and volume fraction (see the supplemental material Figure S4), but found that none produce hyperuniform structure. 

Next we investigated if the long-range order could be generated by the phase separation process.
To model this, we perform two-dimensional numerical simulations of the Cahn-Hilliard equation, which was formulated to describe spinodal decomposition~\cite{cahn1958free}, and has been applied to biomolecular systems~\cite{gasior2020modeling,laghmach2020liquid,mao2019phase}. 
The equation represents a free energy expansion of an inhomogeneous liquid with spatial- and time-dependent concentration c({\bf x},t), and includes the sum of the free energy of a homogeneous liquid with a term that penalizes concentration gradients.
Simulations start from an initial condition where the concentration field is uniformly random $c({\bf x},0)=[-0.5-10^{-4},-0.5+10^{-4}]$ which produces droplets of $c=+1$ in a background of $c=-1$ that occupy 25\% area fraction. 

Our simulations show that the structures formed by a Cahn-Hilliard simulation (Fig.~\ref{fig:NS_CH}d magenta) are hyperuniform with characteristic exponent $\alpha = 4$.
This exponent is consistent with previous calculations of equal-ratio mixtures undergoing Cahn-Hilliard dynamics~\cite{maRandomScalarFields2017}, which don't form droplets but rather percolated patterns.
To test if dimensionality modifies the characteristic exponent, we carried out Cahn-Hilliard simulations in two dimensions and three dimensions, finding $\alpha = 4$ in both cases (see the Supplemental Material). 

While the Cahn-Hilliard simulation produces droplets with hyperuniform scaling, the structures generated do not scale with the same exponent as the experiments: they have much stronger suppression of density fluctuations, $\alpha = 4$, as opposed to the experimental value, $\alpha = 2$. 
We posit that the discrepancy is due to the Brownian motion of the droplets, which occurs in experiment, but is not captured by the simulation.
The difference can be seen by comparing movies of the experiment and the Cahn-Hilliard droplet formation process in the Supplemental Material.
Cahn-Hilliard droplets do not move; they either grow or shrink via Ostwald ripening, so coalescence events only occur when droplets grow large enough to merge.
In contrast, the Brownian motion of the nanostar droplets allows for droplet coalescence without growth. 

Diffusion has also been shown to modulate the hyperuniform scaling in several non-equilibrium numerical models, including absorbing-state models in the active phase~\cite{hexner2017noise}, chiral active particle simulations~\cite{lei2019nonequilibrium}, model soft polymers~\cite{chremos2018hidden} and driven hard-sphere fluids~\cite{lei2019hydrodynamics}, all producing $\alpha=2$.
To explore this, we add an approximation of Brownian motion to the simulation.
Particularly, we locate and size the droplets~\cite{crocker1996methods}, then displace them in a random direction with a magnitude, $\delta$, proportional to each drop's Stokes-Einstein mobility, i.e. $\delta \sim 1/R$, where $R$ is droplet radius.
We find that randomly displaced Cahn-Hilliard droplets recover the $\alpha = 2$ scaling from the experiments (Fig.~\ref{fig:NS_CH}d).
We see quantitative agreement of $\psi(q \rightarrow 0)$ between the experiment and the randomly displaced Cahn-Hilliard drops if we choose a displacement magnitude equal to one average particle spacing (see Supplemental Material Figure S5).
For small displacements, the Cahn-Hilliard hyperuniform scaling $\alpha = 4$ is retained.
As the displacement magnitude approaches the mean droplet size $\psi(q)$ starts to exhibit $\alpha = 2$ at the lowest $q$-values.
For displacements larger than an average droplet size, the structure looks random $\psi(q) \sim q^0$ for length scales smaller than the displacement size while $\psi$ is hyperuniform with $\alpha = 2$ for larger length scales.
This suggests that the nanostar droplets form via a Cahn-Hilliard phase separation, but then diffuse one average particle spacing before being imaged. 
In the experiment, we never observe the $\alpha = 4$ scaling predicted by Cahn-Hilliard, even at the earliest times, indicating that the lateral diffusion occurs concurrently with sedimentation.

\section{Dynamical Assembly of droplet structures}

\begin{figure*}
\includegraphics[width=17cm]{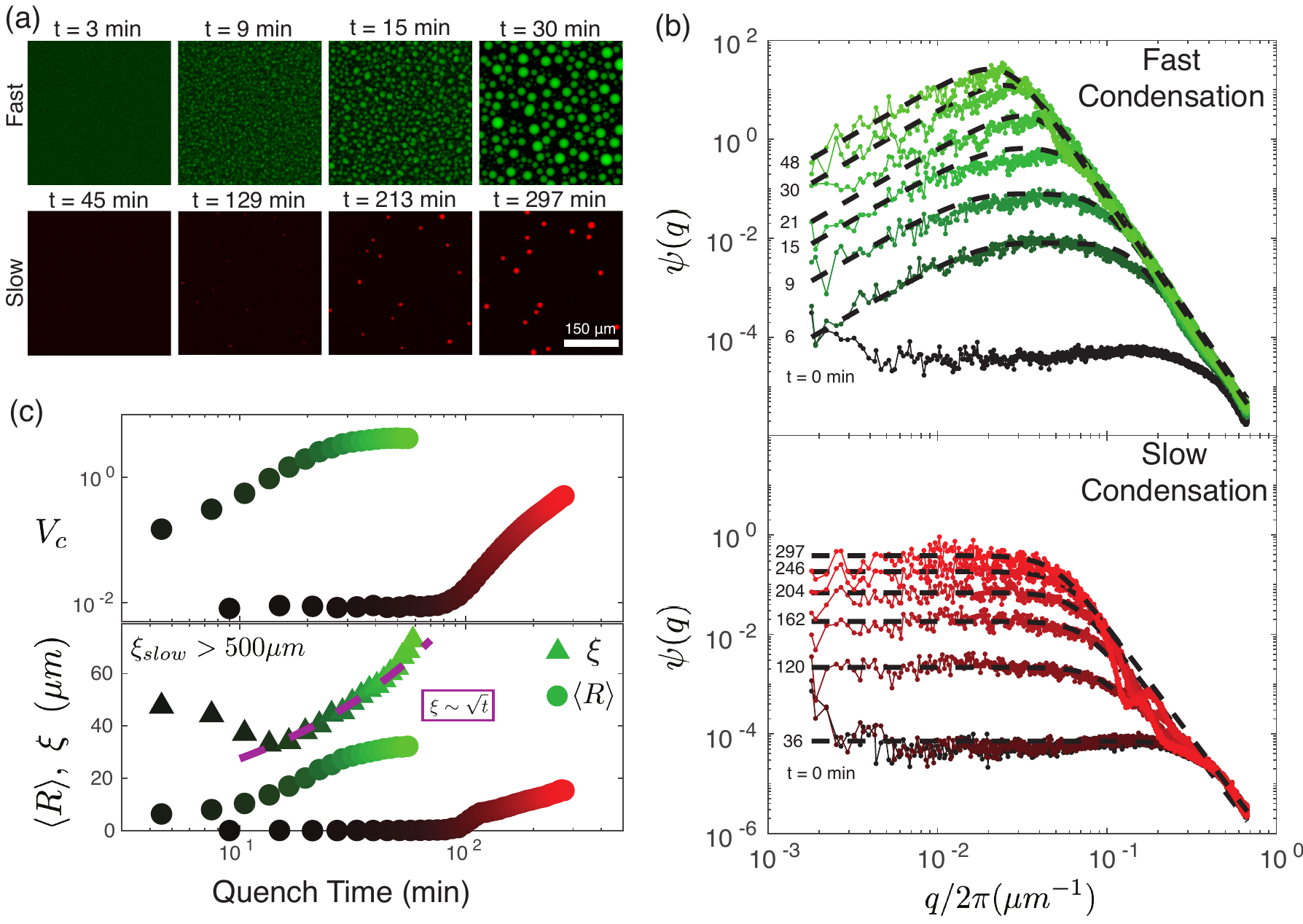}
\caption{\label{fig:Chi_time} Dynamics of hyperuniform structure formation. (a) Images of nanostar droplet formation for a fast condensation process ($30\mu$M nanostars, $500$mM ionic strength) and a slow condensation process ($10\mu$M nanostars, $100$mM ionic strength).
(b) Scattering functions, $\psi(q)$, for the fast- and slow-condensing systems, for different times after quench. 
The scattering function in both cases displays a large-$q$ decay indicative of the particle size as expressed by Eq.~\ref{eq:1}.
At small-$q$, the fast condensation process produces hyperuniform structures, $\psi \sim q^2$, as droplets form, while structures are not hyperuniform for slow condensation.
Each $\psi(q,t)$ is fit to Eq.~\ref{eq:1} (dashed lines).
(c) Fit parameters of Eq.~\ref{eq:1} to $\psi(q)$ data show the volume of condensate, $V_c$, increases quickly and plateaus for the fast process, while slow condensation only produces droplets after a significant delay ($\sim$100 min). 
This same dynamical behavior is reflected in the mean droplet size, $\langle R \rangle$ (circles). 
For fast condensation, the hyperuniformity length scale, $\xi$, (triangles) increases with time, after a brief decline at early times when droplets are still forming.
At later times ($>$10min), $\xi$ displays scaling consistent with diffusive equilibration of the droplet structure, $\xi \sim \sqrt{t}$.
Plotted $\psi$ values are the average of three different quenches and two different sample preparations. 
}
\end{figure*}

The success of the ``Cahn-Hilliard plus diffusion" picture (Fig.~\ref{fig:NS_CH}d) assumes that the timescale for droplet growth is much smaller than the timescale for Brownian motion.
Therefore, it is compelling to explore the density fluctuations in the opposite regime, where Brownian motion occurs on shorter timescales relative to droplet growth.
We found that a sample with a large nanostar concentration ($30\mu$M), and at high ionic strength (KAce = $500$mM), formed droplets much more quickly  than a sample with a lesser concentration ($10\mu$M), and lower ionic strength (KAce = $100$mM) (Fig.~\ref{fig:Chi_time}).
In both cases, the nanostars are prepared in the melted state, then the temperature is quenched to $38^{\circ} $C and held for the length of the experiment.

The vastly different phase separation dynamics produce qualitatively different droplet structures. 
For both cases, when the temperature is above the melting point, unbound nanostars form a homogeneous fluid, and we measure a constant random structure, $\psi(q) \approx 10^{-4}$ (Fig.~\ref{fig:Chi_time}b).
As the system is quenched below melting, we observe the emergence of hyperuniform scaling $\psi(q\rightarrow 0) \sim q^\alpha$ with $\alpha = 2.0 \pm 0.1$ in the case where the droplets form quickly.
The $\alpha = 2$ scaling emerges almost immediately after droplets form, at quench times of about 6 minutes.
We don't observe evidence of the $\alpha = 4$ scaling seen in the Cahn-Hilliard simulations.
In contrast, the random structure of the homogeneous fluid, $\psi(q\rightarrow 0) \sim q^0$, is preserved after quenching the slow system, even just after droplets appear. 
This observation supports the hypothesis that droplet growth kinetics relative to Brownian motion controls the long-range structures.

That we can control the hyperuniformity of the droplet structures by the relative timescales of droplet growth and diffusion implies there is a hyperuniformity length scale, $\xi$.
For length scales shorter than $\xi$ the density fluctuations are random, $\psi \sim q^0$, while for length scales longer than $\xi$ density fluctuations are hyperuniform, with $\psi \sim q^2$ scaling.
Due to the mechanisms for hyperuniform organization we have identified, $\xi$ should evolve with time and the only other mesoscopic length scale in the system, the droplet size.
To assess evolution of $\xi$ quantitatively, we fit $\psi(q)$ to a functional form,
\begin{equation}
    \psi(q) = V_c^2 \frac{1 - e^{-(q\xi)^2}}{1 + (q\langle R\rangle)^\beta}
    \label{eq:1}
\end{equation}
with three fitting parameters: $V_c$, the condensate quantity; the hyperuniformity length scale, $\xi$; and the mean particle size, $\langle R \rangle$.
Eq.~\ref{eq:1}, while phenemonological, is justified by decomposing the scattering function $\psi(q)$ into a structural component $S(q)$ that incorporates long-range structure and a form factor $F(q)$ that accounts for the particle size and profile.
This is done by assuming that the scattering function is separable, such that the small-$q$ behavior is dominated by $S(q)$ and goes to $1$ at large-$q$, and vice-versa for $F(q)$.
To replicate the hyperuniformity we measure in Fig.~\ref{fig:NS_CH}, we chose the form $S(q) = 1 - e^{-(q\xi)^2}$ where $S(q) \sim q^2$ for length scales larger than $\xi$.
This exponential form is the solution to the diffusion equation in Fourier space where $\xi=\sqrt{Dt}$, where D is the diffusion coefficient of the droplets~\cite{fick1855ueber,fourier1822theorie,philibert2005one}.
The choice of form factor, $F(q) = 1/(1+q \langle R\rangle)^\beta$, with $\beta = 5$, is phenomenological, but related to the form factors expected for polydisperse droplets (See Supplemental Material Section III).

For all cases of fast and slow droplet growth, the functional form, Eq.~\ref{eq:1}, fits well over the whole range of $q$-values and times after the quench (Fig.~\ref{fig:Chi_time}a-b).
The fit parameter $V_c(t)$ reflects the qualitative differences in nanostar droplet growth dynamics; the fast condensation process produces droplets quickly, $<5$ minutes, and reaches a plateau value at ~30 minutes, while $V_c(t)$ of slow condensation doesn't show evidence of droplet formation until $\sim 100$ minutes and does not plateau even after 300 minutes (Fig.~\ref{fig:Chi_time}c).
The mean droplet size $\langle R\rangle(t)$ reflects the dynamics of the condensate volume $V_c(t)$, quickly plateauing for fast condensation while a significant delay plus slow droplet growth is characteristic of the slow condensation process (Fig.~\ref{fig:Chi_time}c).

For the slow condensation, we cannot resolve a hyperuniformity length scale $\xi$.
$\xi$ for the fast condensation process displays a non-monotonic behavior as droplets form, at first decreasing then increasing again at longer times.
The early time decrease in $\xi$ could be due to the rapid droplet growth in this phase of the quench, where small, quickly diffusing droplets slow down dramatically as they grow.
In the late phase of droplet growth, $\xi$ is consistent with a diffusive scaling, where modes of a length scale $\xi$ relax back to equilibrium with $\xi \sim t^{1/2}$.
The diffusive behavior of $\xi$ doesn't only agree in the plateau region of $V_c(t)$ but at even earlier times as well.

It is tempting to assign the slow dynamics to a nucleation process, and the fast to spinodal decomposition~\cite{cahn1961spinodal,gibbs1879equilibrium}.
For sufficiently deep quenches, one expects the phase separation is spinodal, defined by dynamics that lack a time delay or an energy barrier; i.e. the mixture spontaneously demixes once the system is quenched below a certain temperature.
However, our optical microscopy method cannot detect the existence of droplets smaller than about $1 \mu$m.
We performed dynamic light scattering (DLS) measurements on the ``slow" condensation process and find that small droplets form soon after the temperature quench (see Supplemental Figure S7), and are small enough to diffuse the entire image size before we detect droplets on the microscope.
Therefore, hyperuniformity only results from droplets that grow quickly enough to diffuse slowly, thus preserving the phase-separated structure before droplet Brownian motion erases the droplet correlations.

\section{Droplet size controls $\xi$}

\begin{figure}
\includegraphics[width=8.5cm]{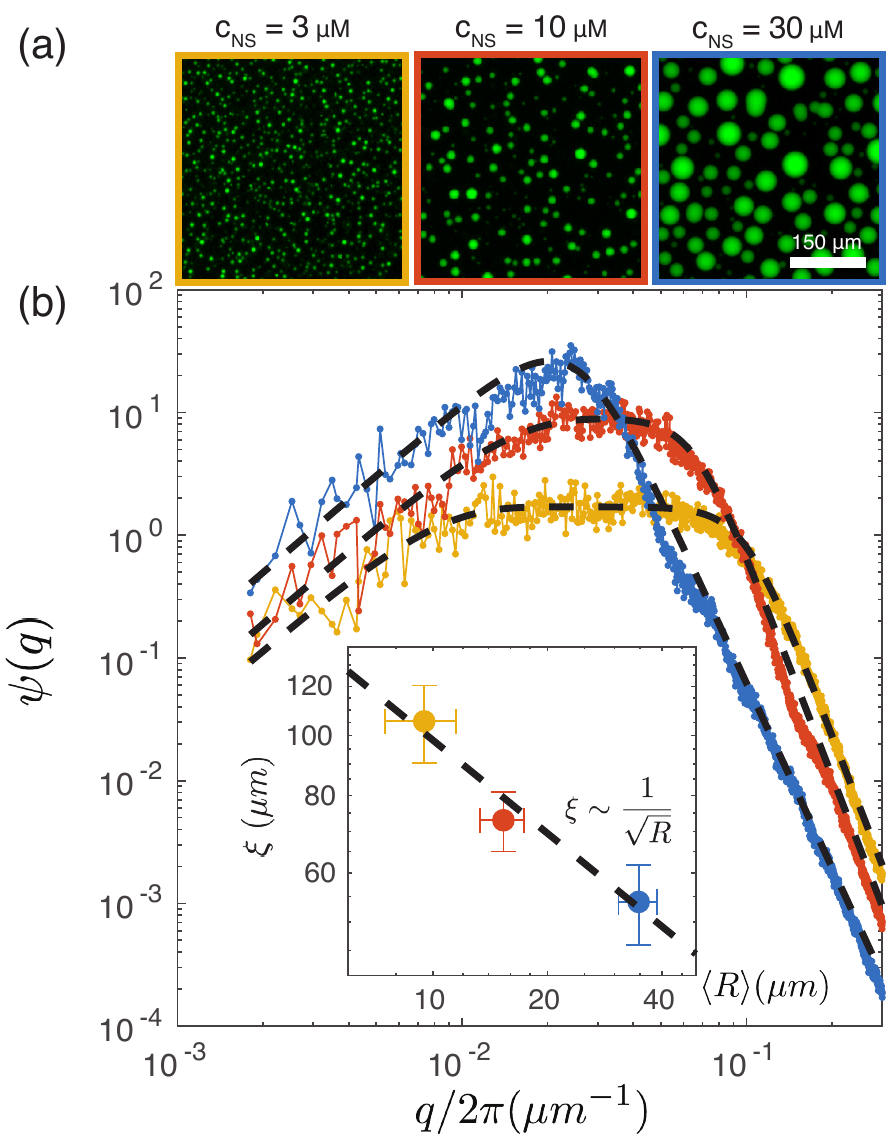}
\caption{\label{fig:Chi_conc} Droplet size influence on hyperuniform structure. (a) Images of nanostar droplet samples of different bulk concentrations, 60 minutes after quench.
(b) For all concentrations, the scattering function displays a large-$q$ decay associated with the particle size as well as a small-$q$ hyperuniform scaling region with $\alpha=2$, but with different length scales $\langle R \rangle$ and $\xi$, leading to a plateau region for smaller droplets. Dashed lines are fits to Eq.~\ref{eq:1}.
Inset: The relation between the fit values of $\xi$ and $\langle R \rangle$ is consistent with Stokes-Einstein diffusion, which predicts $\xi \sim 1/\sqrt{\langle R \rangle}$ (dashed line).
Plotted $\psi$ values are the average of three different quenches for each concentration.  
}
\end{figure}

The role diffusion plays in the long-range structure is illuminated by varying the droplet size.
We vary the mean droplet size, from $\langle R \rangle \approx 10 \mu$m to $\approx 40 \mu$m, by adjusting the nanostar concentration (Fig.~\ref{fig:Chi_conc}a), then measure the scattering function a fixed time after quench.
We find a separation of the length scales, $\xi$ and $\langle R \rangle$, as the droplets get smaller, resulting in a plateau in $\psi(q)$ at intermediate $q$-values (Fig.~\ref{fig:Chi_conc}b).
This occurs because, as droplets get smaller, they diffuse further in a given time; thus, as $1/\langle R \rangle$ moves to larger $q$,  $1/\xi$ moves to smaller $q$.
We confirm this quantitatively by fitting the $\psi(q)$ to Eq.~\ref{eq:1},  extracting best-fit $\xi$ and $\langle R \rangle$, and finding $\xi \sim 1/\sqrt{\langle R \rangle}$ (Fig.~\ref{fig:Chi_conc}b Inset), which is consistent with the Stokes-Einstein prediction for spherical droplets, $D \sim 1/R$, given the expectation that $\xi \sim \sqrt{D t}$.
The emergence of the plateau, along with the relation of the length scales, confirms that the hyperuniform structure is erased by diffusion.
That we find the hyperuniformity length scale $\xi$ follows the Stokes-Einstein form $\xi \sim 1/\sqrt{R}$ is one indication that the system is near equilibrium on length scales up to $\xi$.
This suggests that the phase separation process represents a perturbation that then relaxes via equilibrium statistics.

\section{Distinguishable Hyperuniformity of Multicomponent Phase separation}

\begin{figure}
\includegraphics[width=8.5cm]{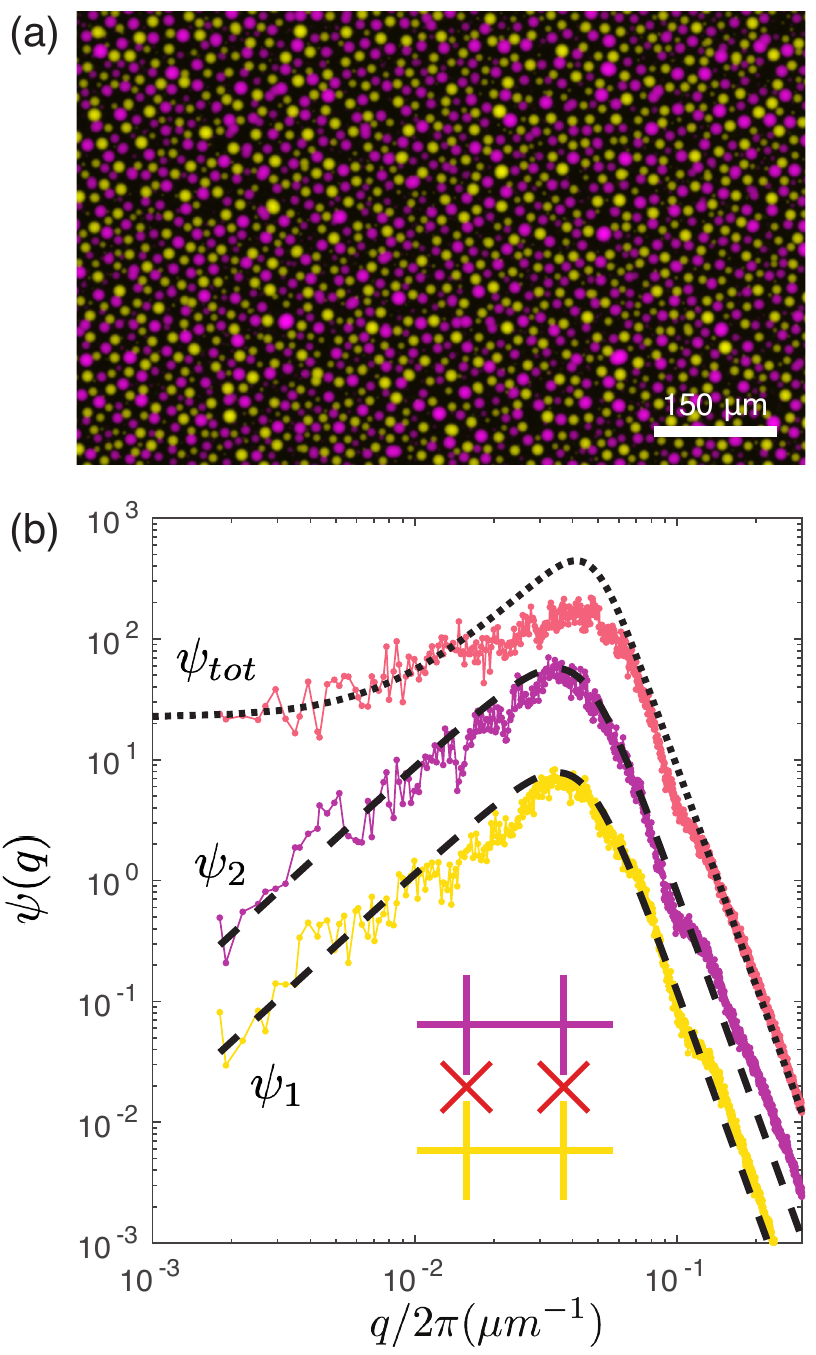}
\caption{\label{fig:Chi_2color} Hyperuniformity of multicomponent nanostar droplets. (a)~Subsection of a two-color fluorescent image of a two-component mixture of nanostars with orthogonal sticky-ends GGAATTCC (yellow), and CTAGCTAG (pink). (b)~The scatting function shows that each component ($\psi_{1}, \psi_{2}$) retains  $\alpha=2$ hyperuniform scaling (dashed line) even in the presence of another orthogonal nanostar species. However, the scattering function of both droplet species ($\psi_{tot}$) is not hyperuniform, but rather consistent with the scattering function of a 29\% area fraction hard disk fluid (dotted line) due to excluded-volume interactions between unlike droplets.
Plotted $\psi$ values are the average of five different quenches. 
}
\end{figure}

The ability to precisely control nanostar sequences allows us to test the phase-separation driven hyperuniformity hypothesis through multicomponent condensation.
We design two orthogonal nanostar species such that each species' sticky ends only bind to other particles of the same species~\cite{jeonSequenceControlledAdhesionMicroemulsification2020,gong2022computational}. 
Particularly, we create nanostars with two binding sequences, GGAATTCC and CTAGCTAG, which are palindromic sequences with similar binding strength (ensuring strong same-species binding, and similar melting temperatures), but that are chosen to minimize the cross-species hybridization energy.
Accordingly, when quenching solutions of the two nanostar species, we observe two separated droplet species, without adhesion (Fig.~\ref{fig:Chi_2color}a).
We find the structures of each droplet species, individually, i.e. $\psi_{1}$ and $\psi_{2}$, are hyperuniform (Fig.~\ref{fig:Chi_2color}b) with the same $\psi(q \rightarrow 0) \sim q^2$ scaling that we measure in the single-species case (Fig.~\ref{fig:NS_CH}d).
However, when both species are analyzed together, the total scattering function $\psi_{tot}$ is not hyperuniform (Fig.~\ref{fig:Chi_2color}b).
Instead, the structure is consistent with that expected for a hard disk fluid occupying area fraction $\phi=0.29$ (see Methods for more details), which is consistent with a significant role played by excluded volume interactions between the two orthogonal droplets. 
There is a small deviation between the hard-disk prediction and the measured $\psi_{tot}$ near the peak (Fig.~\ref{fig:Chi_2color}b); this likely occurs because the hard disk solution assumes a monodisperse population of particles, whereas the droplets are polydisperse.

The lack of hyperuniformity in the summed (two-species) structure is opposite to that found for multi-species cones in the chicken eye retina~\cite{jiaoAvianPhotoreceptorPatterns2014}.
While the mechanism at play in the chicken retina is unclear, here we attribute the lack of multi-species hyperuniformity to the independence of the phase-separation processes in each species.
We have shown that the spatial structure associated with Cahn-Hilliard spinodal decomposition dominates the nanostar system, leading to hyperuniformity within each species individually.
However, the two phase separation processes are made independent by the molecular design, leading to random cross-species droplet placements, and a non-hyperuniform total structure ($\psi_{tot}(q \rightarrow 0) \sim q^0$).
This result also suggests, very clearly, that sedimentation is not the mechanism for long-range organization, because droplets from the two species do interact hydrodynamically, but do not show mutual hyperuniformity.  

\section{Hyperuniform structures formed by electrostatic coacervation}

The hypothesis that long-range ordering of nanostar droplets is driven by spinodal decomposition and Brownian motion suggests that  $\alpha = 2$ hyperuniformity should be observed in many phase-separating soft matter systems, not limited to DNA nanostars.
To test this hypothesis, we perform experiments on droplets formed by the electrostatic coacervation of positively-charged polymers (polylysine; PLL) and negatively charged small molecules (the nucleotide ATP) ~\cite{lu2021temperature}. 
For a fair comparison to the nanostar droplet system, 
we work in solution conditions that induce the formation of droplets of similar size.
The sample is prepared in a PCR tube, heated to $60^\circ$C to evenly mix, then loaded into a polyacrylamide-coated glass capillary at room temperature to induce phase separation.

\begin{figure}
\includegraphics[width=8.5cm]{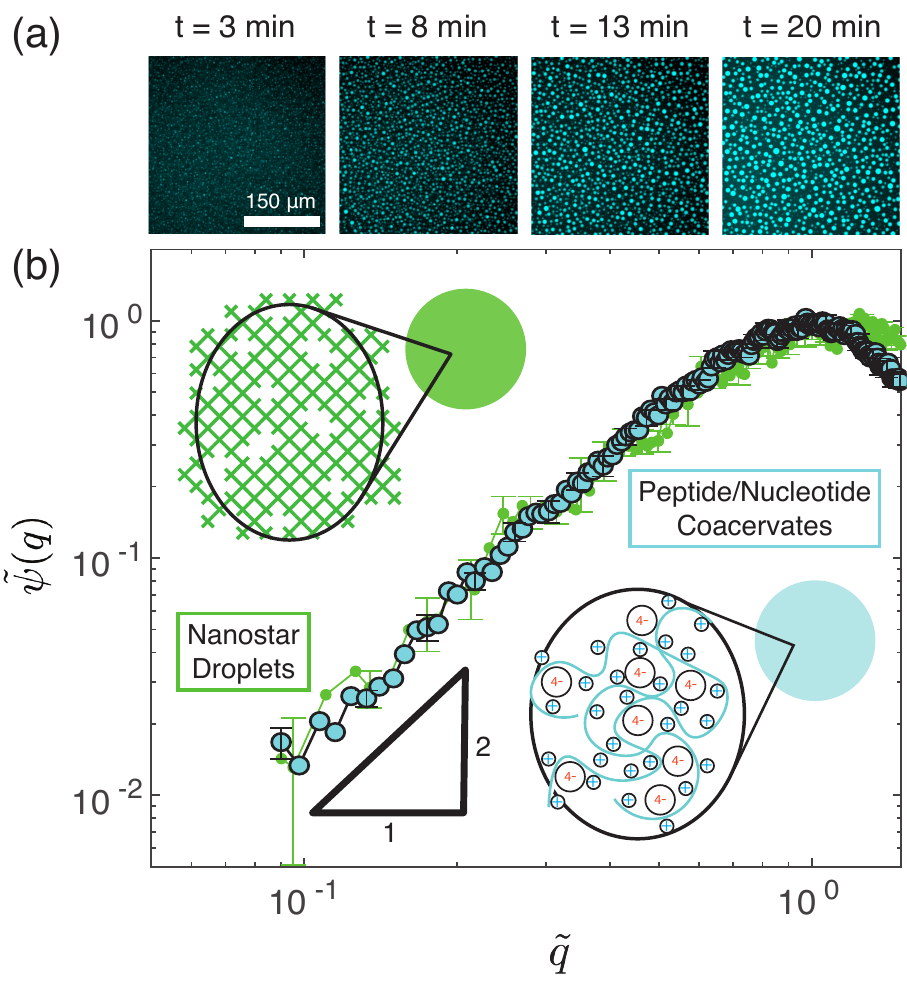}
\caption{\label{fig:PLL} 
Electrostatic coacervates form identical hyperuniform structures.
(a) Images of electrostatic coacervate formation for a system of polylysine (100-mer, $40$ mM monomers) and ATP ($40$ mM) in $400$ mM KAce.
Fluorescein-UTP is added at 10 nM for visualization.
(b) The scattering function, $\psi(q)$, for the polylysine/ATP droplets (cyan), 20 minutes after phase separation, displays $\alpha = 2$ hyperuniform structure that quantitatively agrees with the scattering function of nanostar droplets (green) reproduced from Fig.~\ref{fig:NS_CH}d. 
$\psi$ is averaged over 10 distinct locations in the sample.
Values of $\psi$ are normalized for comparison, $\tilde{\psi}=\psi/\psi_{peak}$ and $\tilde{q}=q/q_{peak}$.
}
\end{figure}

We calculate $\psi$ for the PLL/ATP droplets twenty minutes after phase separation (Fig.~\ref{fig:PLL}). 
We find that the droplet patterns formed by electrostatic coacervation display $\alpha = 2$ hyperuniform structures, and agree quantitatively with the hyperuniformity of nanostar droplets.
This agreement suggests that the spatial organization of phase separated droplets is not limited to the unique characteristics of DNA nanostars but rather is a general feature of phase-separating systems that experience Brownian motion.

\section{Conclusions}
We observe assemblies of DNA nanostar condensates and electrostatic polylysine/ATP coacervates that exhibit strongly suppressed, hyperuniform density fluctuations that result from the interplay of a spinodal decomposition phase separation process and Brownian motion. 
We demonstrate control of the spatial structure via a diffusive length scale $\xi$ which produces $\psi \sim q^2$ hyperuniformity for length scales larger than $\xi \sim \sqrt{t/\langle R\rangle}$, for quench time $t$ and mean droplet size $\langle R\rangle$.
The bottom-up design of orthogonal droplet species preserves the long-range hyperuniform order of individual components, while the hyperuniformity of the entire assembly is destroyed.
These observations further support that phase separation, as driven by specific biochemical interactions, results in the measured long-range order, rather than any non-specific physical mechanism, such as droplet-droplet hydrodynamic interactions.
Hyperuniformity induced by phase-separation is also a general feature, as both DNA nanostar droplets and electrostatic PLL/ATP coacervates share the  hyperuniformity exponent $\alpha = 2$ characteristic of droplet Brownian motion.

The suppression of density fluctuations of phase-separated droplets is remarkably strong. 
Only one previous colloidal-scale experiment of periodically sheared suspensions~\cite{wilkenHyperuniformStructuresFormed2020b} has demonstrated control of the hyperuniform structure in three dimensions with $\alpha = 0.25$, which is much weaker than the phase-separated droplets $\alpha=2$.
In addition, sheared suspensions are only  hyperuniform at the critical point of a dynamical phase transition and display random structures on either side of the transition.
In contrast, the hyperuniformity of phase-separated droplets occurs at a range of salts, temperatures and nanostar concentrations.
Finally, the structures of sheared suspensions are locally anisotropic due to the shearing, which limits their application to band gap materials, while phase-separated droplets are isotropic on all scales (Supplement Figure S1).


The mechanistic link between phase separation and the emergence of hyperuniformity can be understood by considering the Cahn-Hilliard model.
Fourier analysis of the Cahn-Hilliard equation indicates the amplitude $A_q(t)$ of a concentration mode with wavevector $q$ evolves from a homogeneous initial state where $A_q(0) \sim \text{constant}$ with the form $A_q(t) \sim e^{ D_{NS}(q^2 - \gamma q^4) t}$ at early times~\cite{elliott1989cahn}.
$D_{NS}$ is the particle (nanostar) diffusion coefficient, and $\gamma$ is related to the interfacial energy between the two phases. 
There is thus a fastest growing mode (with $q_{max} \approx 1/\sqrt{\gamma}$) that separates low $q$, diffusion-dominated, slow-growing modes, from high $q$, interface-dominated, exponentially-suppressed modes.
$A_{q}$ grows with time for all $q<q_{max}$, indicating phase separation increases density fluctuations on all long length scales.
However, at increasingly small $q$ the amplitude grows increasingly slowly (i.e. to lowest order, as $A_q(t) \approx D_{NS}q^2 t$); this is because diffusion is progressively slower over longer scales. 
Thus, at finite time the density fluctuations at low $q$ are much smaller than those at $q_{max}$, and it is this relative suppression that corresponds to the hyperuniform structure. 
Ultimately, then, the long length scale structure of the nanostar droplets arises from two diffusion mechanisms: the diffusion of individual nanostars drives the hyperuniform Cahn-Hilliard mode structure $(S_{CH}(q) \approx (D_{NS}t)^2q^4)$, while, at long times, droplet diffusion erodes the hyperuniformity at progressively longer length scales with the form $S(q) = (S_{CH}(q)-1)e^{-q^2Dt}+1 \approx Dq^2t$.

The hyperuniform structure of droplets demonstrated in this work could lend itself to a variety of applications, in part because droplet diffusion can be arrested, e.g. by adhering droplets to the surface, thus creating near-permanent hyperuniform structures (see Supplemental Material Figure S6).
Given previously-established connections between spinodal decomposition and material properties~\cite{hsieh2019mechanical}, we posit that hyperuniform, multi-component phase-separated DNA could be used as a means to create patterned mechanical metamaterials, because mechanical properties can be programmed via nanostar binding strength~\cite{conrad2019increasing}.

Disordered hyperuniform structures may be particularly suited for use as photonic materials.
Hyperuniform structures have been shown to display photonic band gaps in two dimensions, even at remarkably low refractive index contrast ($\Delta n \approx 1.6$)~\cite{man2013photonic}.
Therefore, the hyperuniformity we observe in micron-scale phase-separated droplets may lend itself to the fabrication of bottom-up, self-organized structures with exotic optical properties at visible wavelengths.
While neither DNA nanostar droplets ($\Delta n \approx 0.02$~\cite{jeon2018salt}) or PLL/ATP coacervates ($\Delta n \approx 0.1$~\cite{mccall2020quantitative}) possess the necessary refractive index contrast for band gaps, both materials could be readily modified, e.g. by binding titania nanoparticles to the DNA~\cite{zhang2014adsorption} or PLL/ATP as a scaffold, or by implementing high-refractive index polymers as the phase-separating material itself~\cite{naheed2021synthesis}.
Further, even without large refractive index contrasts, interesting photonic properties might be observed using hyperuniformity.
For example, such materials can display structural color even with small refractive index contrasts; and, indeed, this mechanism is exploited by iridophores that manipulate visible light by altering cellular microstructure using biopolymer components with small $\Delta n$~\cite{land1972physics,sicher2021structural}. 

Alternatively, droplet hyperuniformity might be exploited in chemical reaction schemes analogous to those present in biomolecular condensates~\cite{banani2017biomolecular,henninger2021rna}.
Consider an array of point sources that are emitting a diffusing material.
If those sources are structured with $S(q) \sim q^\alpha$, the fraction of space unoccupied by diffusive material after time $t$, $f_V (t)$ decays as $f_V \sim t^{-(d+\alpha)/2}$, for spatial dimension $d$ \cite{torquato2021diffusion}.
Thus, in $d=2$, material diffusing outward from randomly-distributed sources ($\alpha = 0$) fills space as $f_V \sim t^{-1}$, while material diffusing outward from spatially correlated sources (e.g.~DNA droplets, $\alpha = 2$) fills space much faster, with $f_V \sim t^{-2}$.
In a scheme where the droplets act as sinks, carrying out reactions on an initially-uniform concentration of substrate, the results just quoted indicate that the substrate would need to diffuse further, on average, to reach a droplet/reaction center if the droplets are randomly distributed instead of spatially correlated (e.g. with hyperuniform structure). 
In multi-step reactions (e.g.~involving multiple orthogonal droplet species, as in Fig. 4), this disparity in reaction rate between structured and random droplet arrangements is increased (Supplementary Figure S8).
Thus, hyperuniform droplet patterning could provide a pathway for efficient biochemical reactions when mechanical mixing is not possible.
 
Additionally, the hyperuniformity that we observe here opens new avenues to utilize droplet structural features as a probe for active/non-equilibrium systems.
We have shown that the hyperuniform structure results from a phase-separation process that is then modulated by thermal droplet diffusion governed by equilibrium (i.e. Stokes-Einstein) processes.
Without a simple connection between fluctuations and dissipation in active systems, the droplet size impact on the hyperuniformity length scale $\xi$ would be a viable probe of complex dynamic processes (i.e. an out-of-equilibrium Stokes-Einstein relation).
In addition, recent theoretical work has shown that the hyperuniformity exponent $\alpha$ emerges through a competition between fluctuations and dissipation in generalized Langevin equations~\cite{hexner2017noise}.
This suggests that the long range structure encodes two independent types of non-equilibrium behavior, correlation length $\xi$ and hyperuniformity exponent $\alpha$, applicable to a wide variety of driven systems (e.g. in active systems~\cite{Tayar2022Controlling,testa2021sustained}, or even living matter~\cite{leo2022treatment}).

That disordered hyperuniformity can be driven by phase separation, as observed here, implies that long-range ordering might be more universal than previously considered.
The Cahn-Hilliard equation is pervasive in biology, physics, chemistry, ecology, image processing, and astronomy, implying that hyperuniformity could be broadly observed, from the pattern formation of mussels~\cite{liu2013phase} to the structure of Saturn's rings~\cite{tremaine2003origin}.
Similarly, the modulation of these structures by Brownian motion is not limited to the micro-scale, because random fluctuations exist at all length scales~\cite{murray2006brownian,de2014superdiffusion}.

\begin{acknowledgments}

This work was supported by the W.M. Keck Foundation. We thank Hubert Gao for help with DLS measurements and Anna Nguyen for supplying polylysine.

\end{acknowledgments}

\section{Methods}

\subsection{Sample Preparation}
The sample chamber consists of a borosilicate glass microcapillary tube (Vitrocom) with interior dimensions: $300\mu $m height, $3$mm width and $25$mm length.
Capillaries are internally coated with polacrylamide \cite{sanchez2013engineering} to keep droplets from sticking.
We image the droplets with a Nikon Ti-2 fluorescent microscope with a 10x objective, typically finding approximately 10,000 droplets at a quench time of 43 minutes for the fast condensation process. 
The full size of the images is $~1200\mu m$ (see Supplemental Movies).

For imaging, 10\% of nanostars are modified by labelling one strand with a fluorescent dye on the 3' end. 
We use Cy5 dyed strands (NS1: GGAATTCC) for most experiments, and include a FAM label for the two-component mixtures in Fig~\ref{fig:Chi_2color}.
All samples are buffered with 10mM Potassium Phosphate (pH $\sim$ 6.9).
Salt, in this case Potassium Acetate, is added to concentrations of 100mM and 500mM.

\subsection{Image Analysis}

The scattering function is $\psi({\bf q}) = \langle \tilde{I}({\bf q}) \tilde{I}(-{\bf q})\rangle$, where $\tilde{I}({\bf q})=FFT(I({\bf r}))$ is the two dimensional discrete Fourier transform of the image $I({\bf r})$.
The images are normalized by dividing the melted, uniform nanostar image due to the spatial inhomogeneity of the incident LED light source.
To account for the lack of periodicity in the experimental images, we apply a Hanning window to suppress edge effects, $\tilde{I}({\bf q})=FFT(I(x,y)W(x,y))$ where $W(x,y)=\frac{1}{4}(1 - \cos(x/L))(1 - \cos(y/L))$ and $L$ is the image size.

\subsection{Cahn-Hilliard Simulations}
The evolution of $c({\bf x},t)$ is described by 
\begin{equation}
\frac{\partial c}{\partial t} = D_{NS} \nabla^2 \mu
\end{equation}
where $\mu = c^3 - c - \gamma \nabla^2 c$ is the chemical potential, $D_{NS}$ is the diffusion constant and $\sqrt{\gamma}$ is the length scale of transition region between the phases. 
We solve the Cahn-Hilliard equation numerically in two dimensions with discrete time step $\Delta t=1$ on a 2000x2000 grid with periodic boundaries. 
All simulations are calculated with parameters $D_{NS} = 0.0125$ and $\gamma = 0.1$ and numerically evaluated for $10^5$ time steps, when the dynamics are self-similar and the structures are fractal (the data collapses after scaling $\psi$ by it's peak location $q_{peak}$ and value $\psi(q_{peak})$).

We start from a biased random initial condition $c({\bf x},t)=0.5-10^{-4}X$ where $X$ is a uniform random number between 0 and 1, which generate droplets of $c=1$ in a background of $c=-1$ with an area fraction of ~25\%.
The scattering function calculation is analogous to the experiments, where $\psi({\bf q}) = \langle \tilde{c}({\bf q}) \tilde{c}(-{\bf q})\rangle$.  
$\tilde{c}({\bf q})$ is the discrete Fourier transform of the concentration field.
Here, we do not apply a window because the system is evaluated with periodic boundary conditions.
We show data for the 2D Cahn-Hilliard equation here because we can resolve smaller $q$-values given a fixed number of grid points, which is imposed by memory constraints, though the hyperuniform scaling exponents are the same in two and three dimensions.

\subsection{Percus-Yevick Scattering Function}
We approximate the structure of the hard disk fluid by solving the Ornstein-Zernike equation with the Percus-Yevick closure relation~\cite{percus1958analysis,rosenfeld1990free,hansen2013theory}. 
Then, the numerical Taylor series expansion as $q\rightarrow0$ results in $S(q\rightarrow 0,\phi=0.29) \approx 0.10 + 0.31q^2 + O(q^4)$.
We retain the same form factor, $F(q) = F_1(q) + F_2(q)$, so the hard-sphere scattering function is $\psi(q) \approx V_c^2(0.1+0.31q^2)/(1 + (q \langle R \rangle)^\beta)$.
Therefore, we fit the summed scattering function $\psi_{tot}$ with one fit parameter $V_c$ and find agreement at small $q$ even without any structural fit parameters.
We determine the area fraction by locating and sizing the particles from images (such as in Fig.~\ref{fig:Chi_2color}a).

\normalem
%

\end{document}